\documentclass[aps,prl,reprint,showpacs,floatfix,superscriptaddress]{revtex4-1}
\usepackage{graphicx}
\usepackage{amsmath,amssymb,amsfonts}
\usepackage[bookmarks=false]{hyperref}
\usepackage[utf8]{inputenc}
\UseRawInputEncoding
\usepackage{xcolor}

\begin{document}

\title{Dynamics of water and ethanol in graphene oxide}
\author{Gobin R. Acharya} 
\affiliation{Department of Physics and Astronomy, Wayne State University, Detroit, Michigan 48201, USA}
\author{Madhusudan Tyagi} 
\affiliation{NIST Center for Neutron Research, National Institute of Standards and Technology, Gaithersburg, Maryland 20899-6102, USA}
\affiliation{Department of Materials Science and Engineering, University of Maryland, College Park, MD 20742 USA}
\author{Eugene Mamontov}
\affiliation{ Neutron Scattering Division, Oak Ridge National Laboratory, P.O. Box 2008 MS6473, Oak Ridge, Tennessee 37831, USA}

\author{Peter M. Hoffmann}
\affiliation{Department of Physics and Astronomy, Wayne State University, Detroit, Michigan 48201, USA}

\begin{abstract}
We utilized the momentum transfer(Q)-dependence of Quasi-Elastic Neutron Scattering (QENS) to reveal the dynamics of water and ethanol confined in Graphene Oxide (GO) powder or membranes at different temperatures and in different orientations. The dynamics was measured across different length and time scales using several spectrometers. We found reduced diffusivities (up to 30\% in the case of water) and a depression of the transition temperatures. While water showed near Arrhenius behavior with an almost bulk-like activation barrier in a temperature range of 280-310 K, the diffusivity of ethanol showed little temperature dependence. For both water and ethanol, we found evidence for immobile and mobile fractions of the confined liquid. The mobile fraction exhibited jump diffusion, with a jump length consistent with the expected average spacing of hydroxide groups in the GO surfaces. From anisotropy measurements, we found weak anisotropy in diffusion, with the surprising result that diffusion was faster perpendicular to membrane than parallel to it.
\end{abstract}
\maketitle
\section{Introduction}
Water is the natural solvent and major component of living systems, influencing many biological processes. The increasing world population and the production of industrial waste materials are making the availability of drinking water a global concern. Physical, chemical, and biological methods are used to clean water. A physical method such as membrane filtration is used widely in industries to produce clean water. However, membrane filtration only removes particles larger than a few $\mu m$. As an alternative, permeation through nanoscale materials has been suggested to separate contaminants from water \cite{nair2012unimpeded}. For example, it is observed that water can permeate through Graphene Oxide (GO) membranes with thickness less than $1$ $\mu m$, whereas other liquids, vapors, and gases, including helium, are blocked \cite{nair2012unimpeded}.

Graphene Oxide (GO) consists of atomically thin sheets of graphite containing two regions: an oxidized region, comprised of different types of oxygen-containing functional groups on the basal plane or at the edges, and a pristine region, consisting of strongly bonded carbon networks. The oxidized region acts as a spacer that keeps adjacent layers apart from each other. The interplanar distance is determined by the functional groups attached in the oxidized region \cite{joshi2014precise,loh2010graphene}. Together, the adjacent GO sheets act as a 2D capillary and can be filled by molecular layers of water. GO capillaries provide a nearly frictionless flow of water \cite{joshi2014precise}. After hydrating air-dried GO with water, the interplanar spacing changes from 8.2 \AA \, to 14.2   \AA \, \cite{joshi2014precise, raidongia2012nanofluidic}.

Different contaminants present in water have different interactions with the functional groups attached to the GO sheets. As a result, GO can be used for separation of dissolved species and ions in water \cite{sun2012selective,huang2013salt}. The water inside GO capillaries is restricted by the geometry of GO. The water inside restricted geometries about (0.5 nm-100 nm) is often called confined water. 
There are many ways water can be confined, such as water intercalated in layered solid materials like clay, graphene oxide, or water hydrated on the surface of large biomolecules \cite{cerveny2016confined}.  Physical quantities like viscosity, diffusion, ionic mobility, and the rate of chemical reactions are usually calculated by modelling a liquid as a homogeneous continuum. However, when water is confined, continuum models are often not suitable to explain its properties \cite{khan2010dynamic}. The dynamics of confined liquids depend on various factors such as the nature of the confining surface and liquid interaction with the surfaces \cite{khan2010dynamic}. When a liquid is cooled below its melting point, the dynamics of the liquid can be dramatically slower, and its viscosity can increase \cite{ediger2000spatially}. It has been well known that geometrical restriction can prevent crystallization of water in the temperature region of $150$-$240$ K, where water is in a supercooled state and would crystallize (transition to ice state) immediately in the absence of geometric restriction \cite{cerveny2016confined}.

 The dynamics of confined water also depends on the system in which it is confined. Porous materials have their own identity depending on their pore size, topology, and the degree of hydrophilicity/hydrophobicity. Different materials have different interactions with the confined water. These interactions cause system-dependent alterations in the dynamic behavior of the confined water \cite{cerveny2016confined}. These system-dependent alterations instigate dynamic heterogeneities \cite{shekhar2014universal} and are often associated with "non-universal" relaxation behavior \cite{cerveny2016confined}.  This "non-universal" relaxation behavior of confined water is observed on disordered porous substrates such as silica hydrogel \cite{cammarata2003structure}, Vycor glass \cite{zanotti1999relaxational}, molecular sieves \cite{zanotti1999relaxational}, and graphite oxide \cite{cerveny2010dynamics}, among others. However, water intercalated in a regular structures such as certain mesoporous silica substrates, is less influenced by surface interaction and exhibits more "universal" relaxation behavior \cite{grunberg2004hydrogen,liu2005pressure, sattig2013dynamic}.
 
 Ethanol is also an important fluid in chemical, biological, and industrial processes. Ethanol is used to produce fuels such as fuel graded ethanol or diesel-ethanol blends. Ethanol as a commercial bio-fuel has been promoted for its potential environmental benefits. Ethanol can be prepared by various processes such as fermentation or catalytic hydration of ethene in the presence of water vapor. The production of pure ethanol in such processes involves the separation of water and ethanol. Conventionally, water and ethanol are separated by distillation, which is expensive in terms of energy and costs. Distillation consumes more than half of the total energy in the production of ethanol from the fermentation process \cite{nakao1987continuous}. Therefore, the separation of ethanol from water by using GO could replace the conventional distillation process \cite{nair2012unimpeded}.
 
 Experimentally, the microscopic dynamics of confined liquids can be explored by Quasi-Elastic Neutron Scattering (QENS). QENS is an experimental technique which can probe the space-time characteristics of the dynamic processes. QENS can provide information on correlated particle motions in the case of coherent scattering and single-particle motions in the case of incoherent scattering. For hydrogen-rich materials such as water and ethanol, the total scattering cross-section is strongly dominated by self-motions of hydrogen atoms, due to their high incoherent scattering cross-section. The dependence of the relaxation time on the momentum transfer (Q) during the scattering event identifies the physical nature of a dynamic process, e.g., whether it is associated with translation diffusion, localized motion, or rotational motion. However, a single QENS spectrometer is limited by its accessible frequency range. By using several spectrometers such as Back-scattering, time of flight, and Neutron Spin Echo (NSE) one can explore the relaxation times from $10^{-7}-10^{-14}$ s  \cite{cerveny2016confined}. Here, using two spectrometers, BASIS \cite{mamontov2011time} and HFBS  \cite{meyer2003high}, we report the dynamics and the almost isotropic behavior of confined water and ethanol in GO. Our findings might enhance understanding of utilization of GO for the water-ethanol separation applications. 
  
\section{MATERIALS AND METHODS}
Graphene oxide (membranes and powders) was purchased from Advanced Chemical Supplier (ACS Material LLC). To dry the material, it was heated to $ 110 ^o\textrm{C}$ in a vacuum oven for 23 hours. The dried GO powder and membranes were then placed in rectangular sample holders made of aluminum of $5$ cm $\times$ $3$ cm dimensions with an internal thickness  0.5 mm, saturated with water or ethanol vapors, and sealed by indium gaskets for QENS measurement at BASIS at the Spallation Neutron Source (SNS), Oak Ridge National Lab, TN  \cite{mamontov2011time}. 
 
The same samples were packed in an aluminum pouch and sealed in cylindrical cans for fixed-window scans at HFBS, NCNR,  (NIST), Gaithersburg, MD \cite{meyer2003high}. In both cases, the incident neutron beam direction was normal to the side surfaces of the sample holders for the measurements of GO powders. The geometry for the measurements of the oriented GO membranes will be separately discussed in detail. QENS spectra of water and ethanol in GO layers at 10 K (for the sample-specific resolution function), 280 K, 295 K and 310 K were collected at BASIS. The elastic intensity was also measured from 20 K to 310 K with 1 K interval. The incident wavelength band of neutrons was centered at $\lambda$ = 6.4 \AA \, and  Si(111) analyzers Bragg-selected $\lambda$ = 6.27 \AA \ for neutron detection. The data were collected using energy resolution of $3.5$ $\mu$eV  in a dynamic range of $-100$ $\mu$eV to $+100$ $\mu$eV with Q-range of $0.3$ \AA$ ^{-1} $ to 1.9  \AA$ ^{-1}$. The spectra obtained from the dry GO container was subtracted from the spectra obtained from liquid (water and ethanol) intercalated GO container.
Fixed window scans for water in GO were measured at HFBS from 4 K to 300 K with 0.8 K interval. The HFBS spectrometer has an energy resolution of $0.85$ $\mu$eV and covers a $Q$ range of $0.25$ \AA$^{-1}$ to $1.75$ \AA$^{-1}$. Mantid  \cite{arnold2014mantid} and DAVE \cite{azuah2009dave} analysis software were used for the reduction and analysis of the data.

\section{Results and Discussion}
\begin{figure*} 
\begin{center}
\includegraphics[width=18cm]{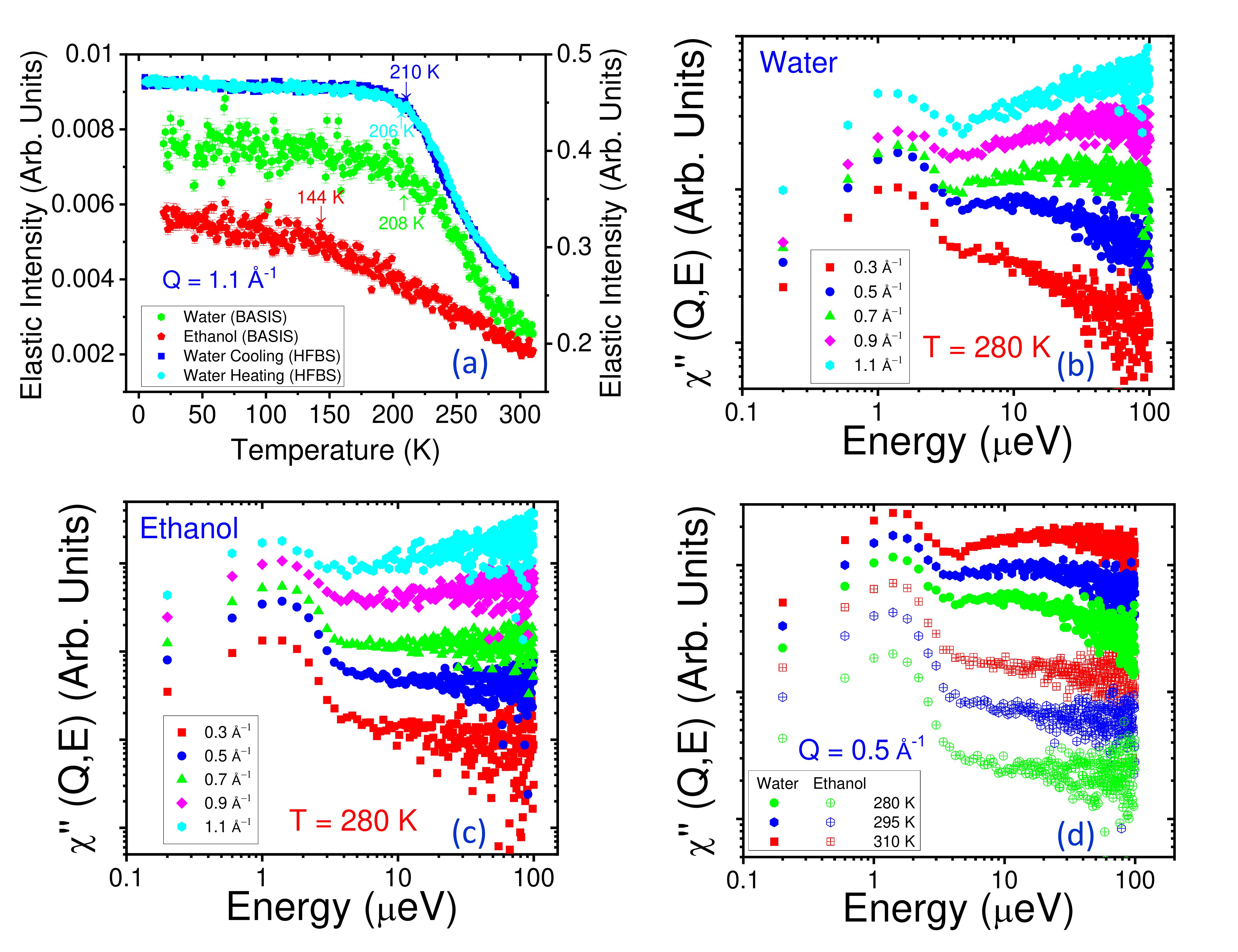}
\caption{(a, upper left) The elastic intensity of water and ethanol intercalated GO measured at BASIS (left vertical axis) and HFBS (right vertical axis). The intensity was measured at BASIS at  $Q = 1.1$ \AA $^{-1}$ from temperature $20$ K to $310$ K with $1$ K interval. At HFBS, the presented intensity was measured  in fixed window scan mode from $4$ K to $300$K with  $0.8$ K  interval. Arrows indicate a "knee" or vertex in the curve. (b, upper right) Dynamic susceptibilities for water at different Q's at 280 K. (c, lower left) Dynamic susceptibilities for ethanol at different Q's at 280 K. (d, lower right) Dynamic susceptibilities comparison of water and ethanol at different temperatures at $Q=0.5$ \AA$^{-1}$.
Please note that in Figures (b), (c) and (d) an arbitrary offset was applied to the individual graphs to separate them for better visibility. }  
\label{fig_1}
\end{center}
\end{figure*}

\begin{figure*} 
\begin{center}
\includegraphics[width=18cm]{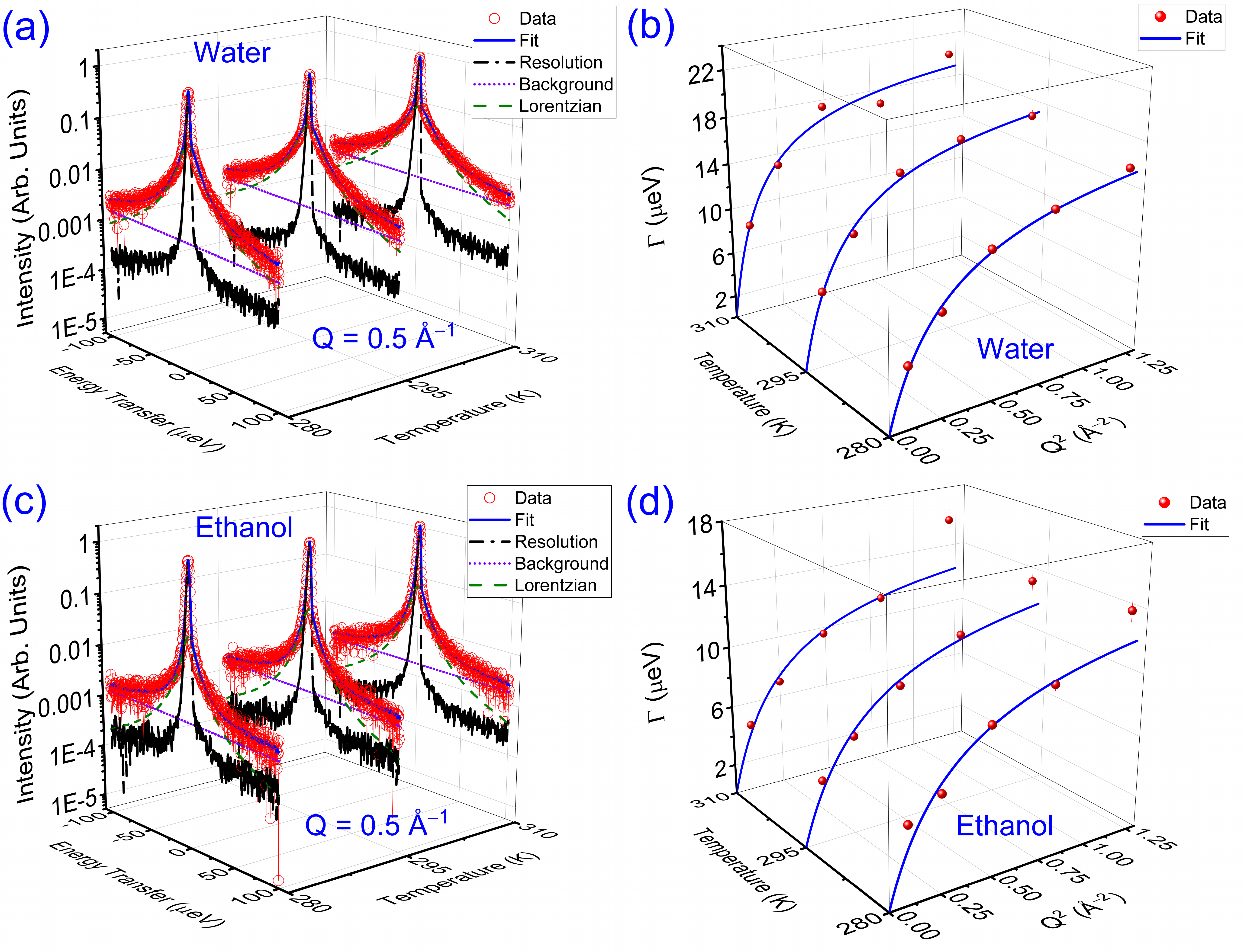}
\caption{QENS spectra measured at BASIS for water (a) and ethanol (c) intercalated in vacuum-annealed GO samples at different temperatures at $Q=0.5$ \AA $^{-1}$.  The symbols represent data, the solid lines represent the model fit, the black dashed dotted lines represent instrument resolution measured at $10$ K, violet dotted lines represent linear background, and the green dashed lines represent Lorentzian. The symbols in (b) and (d) represent the dependence of  $\Gamma$ ($\mu$eV) on $Q^2 $ (\AA $^{-2}$) for water (b) and ethanol (d), where solid lines represent fits using the jump-diffusion model, from which diffusivities are obtained.}
\label{figure_02}
\end{center}
\end{figure*}
Hydrogen atoms have a much larger incoherent neutron scattering cross-section than other atoms, such as carbon and oxygen. As a result, the QENS spectra of hydrogen-containing materials such as water and ethanol represent mainly the scattering from hydrogen atoms, whereas the contribution from other atoms is insignificant. Thus, the QENS spectra measured in our experiment describe the dynamics of hydrogen atoms present in the water and ethanol molecules intercalated in GO.

At first, we performed elastic scattering scans from 20 K to 310 K at BASIS (upon heating) and 4 K to 300 K at HFBS (heating and cooling). The measured elastic intensities are represented in Figure \ref{fig_1}(a). The water measurements did not show a melting or freezing transition around 273 K \cite{osti2016effect}. However, the presence of clear changes in slope (vertices) in the elastic intensity graphs indicate dynamic transitions at lower temperature. The large elastic intensities at temperatures below the vertex points (indicated by small arrows in the plot) reveal that water and ethanol molecules become immobile at low temperature on the time/energy scale probed by the spectrometer, whereas molecules are mobile at temperatures above the vertex points. Therefore there appears to be a transition around 208-210 K for confined water and around 144 K for confined ethanol.

Alternatively, the QENS spectra can be analyzed through their representation as dynamic susceptibilities \cite{ngai2011relaxation} as shown in figure \ref{fig_1}(b). Dynamic susceptibility($\chi'')$ is related to the Bose occupation number, $n_B(E,T)$, as $\chi'' \propto \frac{I(Q,E)}{n_B(E,T)}$ where $n_B(E,T) \propto \left(e^{\frac{E}{k_BT}}-1\right)^{-1}$. Around room temperature the Bose population factor approaches the Maxwell-Boltzmann population factor \cite{kardar2007statistical} and hence $\chi''(Q,E)= \frac{I(Q,E)\cdot E}{K_B T}$ can be obtained from the QENS spectra. The number of peaks in the dynamic susceptibility representation reveals the number of measurable relaxation process that can be detected by a certain instrument in its energy domain. The dynamic susceptibilities shown in figure \ref{fig_1} (b) are therefore helpful in validating any realistic representation of the dynamics. The most representative values are obtained when the elastic peak (broadened by instrument resolution) and a signal associated with a relaxation process are both captured, yet clearly separated. Additionally, the position of the peaks provides information about the time scales of the measured relaxation process \cite{osti2019microscopic}. The observed broad peaks at the low energy end (around 1-2 $\mu$eV) in figure \ref{fig_1}(b)-(d) represent elastic scattering (resolution function of the instrument). The observation of only one additional peak beyond the resolution peak (in the 10-100 $\mu$eV region) indicates there is only one measurable relaxation process within the resolution of the instrument. 

The QENS data [Figure \ref{figure_02} (a,c)] were fitted using the following equation:
\begin{eqnarray}
    I(Q,E) = \Big[ A(Q)\delta(E) + \big(1-A(Q)\big)S(Q,E) \nonumber \\ 
    + B(Q,E)\Big] \otimes R(Q,E) \label{eqn_fitting}
\end{eqnarray}
    
where $A(Q)$ is the weight of the elastic scattering represented by a delta function $\delta(E)$, $S(Q, E)$ is the dynamic structure factor, $R(Q, E)$ is the resolution function of the instrument and $B(Q, E)$ is the background contribution to the measured scattering. As the dynamic susceptibility data indicate, we deal with a single relaxation process. We therefore fitted $S(Q, E)$ using a single Lorentzian function, where $\Gamma(Q)$ represents the half-width at half maximum (HWHM):
 \begin{equation}\label{eqn_366}
    S\big({Q},E \big) = \frac{1}{\pi}\frac{\Gamma(Q)}{E^2+\Gamma^2(Q)}
 \end{equation}

The broadening of QENS spectra beyond the elastic peak, as seen in Figure \ref{figure_02}, is related to the dynamics of the confined molecules. The solid lines represent the fits which give us $\Gamma(Q)$. The graphs in Figure \ref{figure_02} (b and d) show the non-linear dependence of  $\Gamma(Q)$ with $Q^2$. $\Gamma(Q)$ at first increases and then plateaus at higher Q values, which indicates a translational, but non-Fickian, diffusion process, such as jump-diffusion. Jump diffusion models are suitable for molecules which remain in a certain volume for a time $\tau$ before they jump a distance ($l$) to another position. The solid lines represent the fits using the jump-diffusion model governed by the following equation, from which the diffusion coefficient (D) is obtained.
\begin{equation}
     \Gamma(Q) = \frac{\hbar D Q^2}{1+DQ^2\tau_0}
     \label{HWHM}
\end{equation}
Here, $\tau_0$ represents residence time, and D and $\tau_0$ are related by $l^2=6D\tau_0$ from which the jump distance can be obtained. 

\begin{table*} 
\caption{\label{table1}%
    { Parameters obtained for water and ethanol intercalated in Graphene Oxide (powder) for each measured temperature.}}
\begin{ruledtabular}
\begin{tabular}{c c c c c c c c c}
\textrm{Liquid }&
\textrm{T (K) }&
 \textrm{D $\times$ 10$^{-10}$ m$^{-2}$s$^{-1}$}&
\textrm{ $\tau_0$ (ps)}&
\textrm{$l$ (\AA)}&
\textrm{$f$}&
\textrm{$p_0$}&
\textrm{ $a$ (\AA)}
\\
\colrule
         &$280$ & $10.85\pm1.08$ &$34.6\pm2.5 $&$4.7\pm0.3$ & $0.65\pm0.05$ & $0.35\pm0.07$ &$4.4\pm0.8 $\\
Water    &$295$ & $15.21\pm1.26$ &$32.3\pm1.8 $&$5.4\pm0.3$ &$0.59\pm0.06$&$0.40\pm0.10$ &$4.4\pm1.1 $  \\
         &$310$ & $21.91\pm2.32$ &$31.2\pm1.9 $&$6.4\pm0.4$ & $0.57\pm0.06$&$0.41\pm0.11$ &$4.3\pm1.3 $ \\\hline
         &$280$ &  $7.52\pm2.13$  &$43.3\pm9.0 $&$4.4\pm0.8$& $0.88\pm0.01$&  $0.16\pm0.02$  &$3.5\pm0.2 $ \\
Ethanol  &$295$ &  $7.31 \pm1.27$ &$43.6\pm6.0 $&$4.4\pm0.5$&$0.82\pm0.02$ & $0.10 \pm0.05$ &$3.1\pm0.3 $ \\
         &$310$ &  $8.93 \pm1.76$ &$44.6\pm6.3 $& $4.9\pm0.6$&$0.78\pm0.03$ & $0.03 \pm0.09$ &$3.0\pm0.5 $ \\
\end{tabular}
\end{ruledtabular}
\end{table*}

\begin{figure}[h]
\begin{center}
\includegraphics[width=9cm]{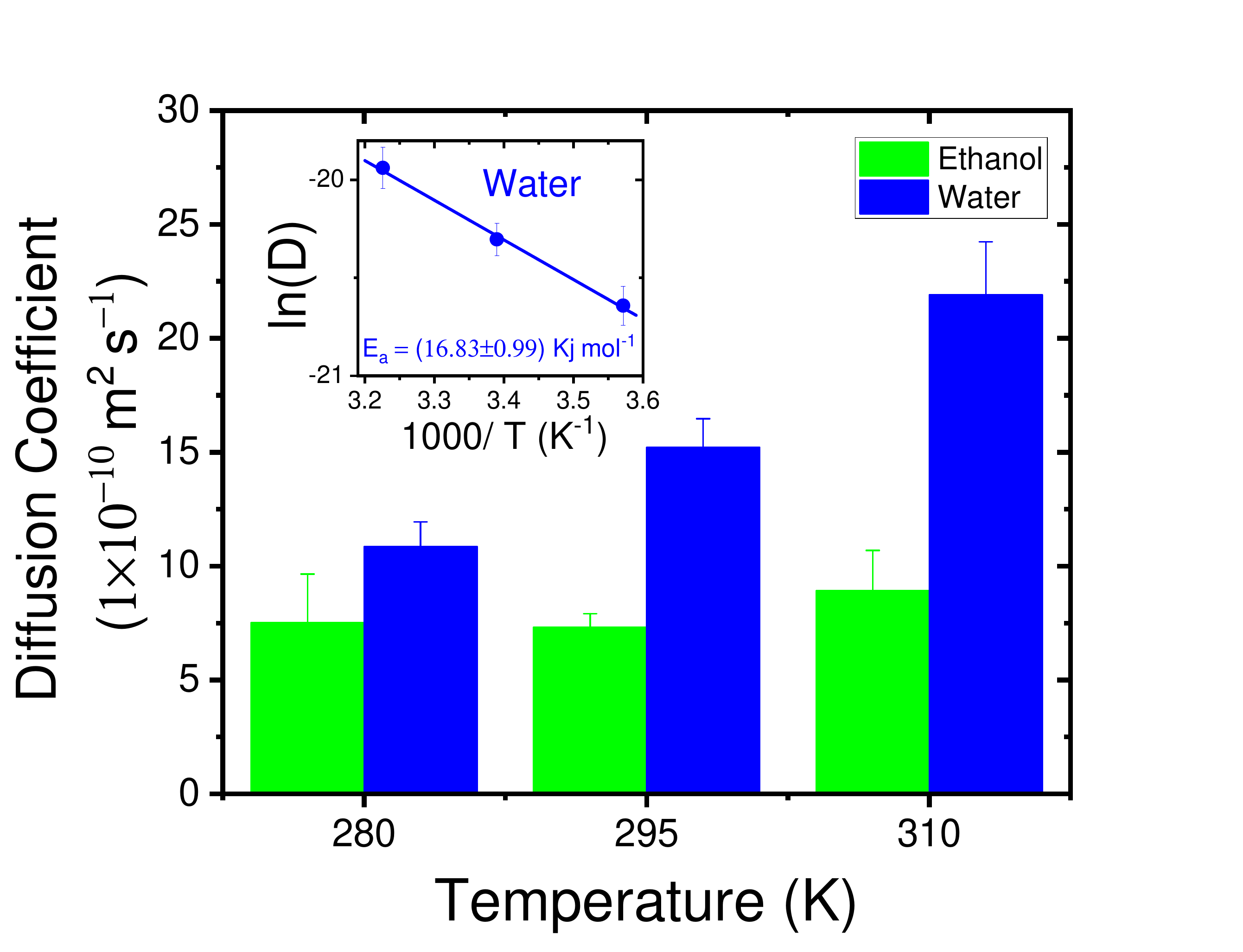}
\caption{The diffusivities of water and ethanol in GO (preliminary vacuum-annealed) at different temperatures. The bars represent the diffusivities of water and ethanol measured at BASIS. The inset graph represents Arrhenius temperature dependence of diffusivity of water in GO measured at BASIS with a linear fit represented by a solid line, from which the activation energy is obtained.}
\label{figure_03}
\end{center}
\end{figure}

In water intercalated in GO, we observed an increase in broadening of the QENS spectra with increasing temperature, indicating activated dynamics. However, we did not observe a significant temperature-dependent increase in broadening for ethanol intercalated in GO. 

Table \ref{table1} summarizes the fitting values obtained from fitting the QENS spectra and applying equation \ref{HWHM} the $Q$ dependence of the HWHM. The measured diffusion coefficient of water and ethanol in GO are also shown in Figure \ref{figure_03}. The measured diffusivities of water and ethanol confined in GO are less than their bulk values \cite{teixeira1985experimental,ben2013influence,kyriakos2016solvent,qvist2011structural,jannelli1996transport}. The diffusivity of confined water in GO is comparable to water confined in substrates such as MCM-41 \cite{takahara1999neutron}, Rutile \cite{mamontov2007dynamics,mamontov2008dynamics},  Mxene \cite{osti2016effect}, Carbon nanotubes \cite{mamontov2006dynamics}, Graphite Oxide \cite{buchsteiner2006water}, ionic polymer \cite{osti2016water}, and others. We observed about a 30\% decrease in the diffusivity of confined water compared to bulk water. 

The diffusivity of water depends on many factors such as level of hydration, hydrophilicity, surface interactions, etc. The dynamic crossover temperature of water ($\approx 208$ K) can be used as indicator of hydrophilicity \cite{chu2007observation} indicating GO as a hydrophilic substrates. The slightly higher diffusivity of water in GO compared to other substrates is in good agreement with the value obtained in other hydrophilic substrates such as 2D  MXene which has a dynamic cross-over temperature at $\approx 200$ K \cite{osti2016effect}.
 
The higher diffusion coefficient of water in GO compared to ethanol can be justified on the basis of  previous studies: The diffusion coefficients of bulk water and methanol overlaps within the uncertainty of measurement \cite{kyriakos2016solvent}. On the other hand, the diffusion coefficient of alcohol decreases in the homologous series when the size of molecule increases \cite{jannelli1996transport}.

The solid lines in the inset graph of Figure \ref{figure_03} represent fits to Arrhenius temperature dependence, assuming $D=D_0\exp{\left(-E_a/k_B T\right)}$. The extracted activation energy $E_a =(16.83\pm0.99)$ kJ$\cdot$mol$^{-1}$ is close to the observed value for bulk water \cite{Krynicki1978H20diffusion,Tsimpa2019H2Odiffusion} This raises the question why the diffusivity under confinement is reduced. To attempt to answer this question, it is worthwhile to have a closer look at the obtained values for $\tau_0$ and $l$ in Table \ref{table1}. The relaxation (residence) times are quite long and beyond what would be expected from single-molecular relaxation times in water or ethanol. To understand this observation, we need to recognize that there is a lower limit of relaxation times that can be measured at BASIS, which is $t_{min}\approx \hbar/\textrm{HWHM}_{max}\approx \hbar/100 \mu \textrm{eV}\approx 7 \textrm{ps}$.   Secondly, it is likely that there are several relaxation mechanisms beyond simple single molecule motion. In confinement, we can expect heterogeneous dynamics, depending on the distance from the confining surfaces \cite{cerveny2016confined} and the presence of epoxide and hydroxyl groups on the graphene surfaces \cite{Lerf1998}. Confinement can also impose collective dynamics, which leads to longer relaxation times \cite{khan2010dynamic, Patil2006}.

It has been found that water diffusion does not fit Arrhenius behavior over extended temperature ranges. Therefore, other fitting methods accounting for the behavior of supercooled water have been used, including Vogel-Fulcher-Tamann (VFT) \cite{chathoth2011QENS} and Speedy-Angell \cite{speedy1976}. With only three data points, it is not possible to make a determination which fits are better. However, fitting to Speedy-Angell $D=D_0\left(T/T_S-1\right)^\gamma$, we found a small decrease in $T_S$, consistent with depressions of phase transition temperatures in confined geometries. The fit yielded $T_S=209.5 $K (bulk: $215$ K \cite{holz2000NMR}), which matches the measured inflection point temperature in elastic intensity (Figure \ref{fig_1}). Fitting to VFT, $D=D_0 \exp\left(B/(T_{S}-T)\right)$, we found an even more depressed value of $T_{S}=196K$.

In our measurements, we subtracted the scattering amplitude of a dehydrated sample from all samples that contained either water or ethanol. Therefore any elastic component would be due to immobilized water or ethanol in the system after re-hydration or adding ethanol. As can be seen in Figure \ref{fig_1}a,b, we have a clear elastic component in the data. Is the immobilized fraction giving rise to an elastic component indicative of strong confinement in this system?

With strong confinement, we would expect the HWHM (Figure \ref{figure_02}) to saturate at low Q values \cite{Bellissent-Funel1995}, but this is not observed. Instead, we observe near free diffusion at low Q values. This is not inconsistent with \cite{Bellissent-Funel1995}, because low-Q saturation would be expected at smaller $Q$ values than probed in Figure \ref{figure_02}( b and d). However, if the molecules were all participating in jump diffusion, with relatively free diffusion at low Q, we would not expect to see a prominent elastic peak after subtracting the data of the dehydrated sample. This suggests that we are probing two species of molecules in the system: An immobilized species, confined to a radius of $a$ (elastic peak), and a mobile species with an average jump distance of $l$ (seen in the HWHM data). The presence of mobile and immobilized species is consistent with the current understanding of the dynamics of water in confined systems \cite{buchsteiner2006water}. The single relaxation time seen in the dynamic susceptibility in Figures \ref{fig_1}b-d can therefore be attributed to jump-diffusive motion of the mobile species in the system.

To further explore the idea of two species that are distinguished by their mobility, we determined the elastic incoherent structure factor (EISF). EISF is the elastic fraction of the signal and is due to molecules that are either completely immobilized on the time scale of the instrument (fraction $p_0$), or confined (fraction $1-p_0$), and therefore able to diffuse only within an effective confinement radius $a$. 

For such a situation, the EISF can be written as \cite{bee1988quasielastic,jacobsen2013nano,shrestha2015effects}:
\begin{align}
 \textrm{EISF} &= p_0 + (1-p_0)\Bigg[\frac{3j_1(Qa)}{Qa}\Bigg]^2 \\
 &=p_0+(1-p_0)\Bigg[\frac{3}{Qa}\bigg[\frac{\sin(Qa)}{(Qa)^2}-\frac{\cos(Qa)}{(Qa)}\bigg]\Bigg]^2
 \label{EISF_eqn}
\end{align}

We found that fitting the EISF (or A(Q) in equation \ref{eqn_fitting}) to this equation, we obtained rather poor fits. We suspect that this because the immobilized/confined species only constitutes a fraction $f$ of the system, while a fraction $1-f$ is participating in jump diffusion as seen from our HWHM fits.

Writing equation (\ref{eqn_fitting}) separately for the two species, and taking $I'(Q,E)=A(Q)\delta(E)+(1-A(Q))S(Q,E)$ we can write:
\begin{align}
    I'_b(Q,E)&=A_b(Q)+(1-A_b(Q))S_b(Q,E)\\
    I'_f(Q,E)&=A_f(Q)+(1-A_f(Q))S_f(Q,E)\\
    I'(Q,E)&=f I'_b(Q,E)+(1-f)I'_f(Q,E)
\end{align}

Here, the subscripts '$b$' and '$f$' stand for bound and free. For the free species, we can assume that $A_f=0$. Moreover, we expect that $\int S(Q,E) dE = \textrm{const.}=1-f$, i.e. the area of $S(Q,E)$ does not depend on $Q$ and only on the mobile fraction. For the bound/confined species, the motion of the locally confined/immobile species can be expected to be fast, leading to a very broad Lorentzian, which is likely not measurable within the energy resolution of the instrument. We therefore take $S_b(Q,E)\approx0$, as the corresponding signal is most likely absorbed into the background $B(Q,E)$.

With this, we can write for the EISF:
\begin{align}
\textrm{EISF}&=\frac{A(Q)}{\int \left(A(Q)+S(Q,E)\right) dE}\\
&=\frac{f A_b(Q)}{f A_b(Q)+(1-f)}
\end{align}

\begin{figure}[h]
    \begin{center}
    \includegraphics[width=9.0cm]{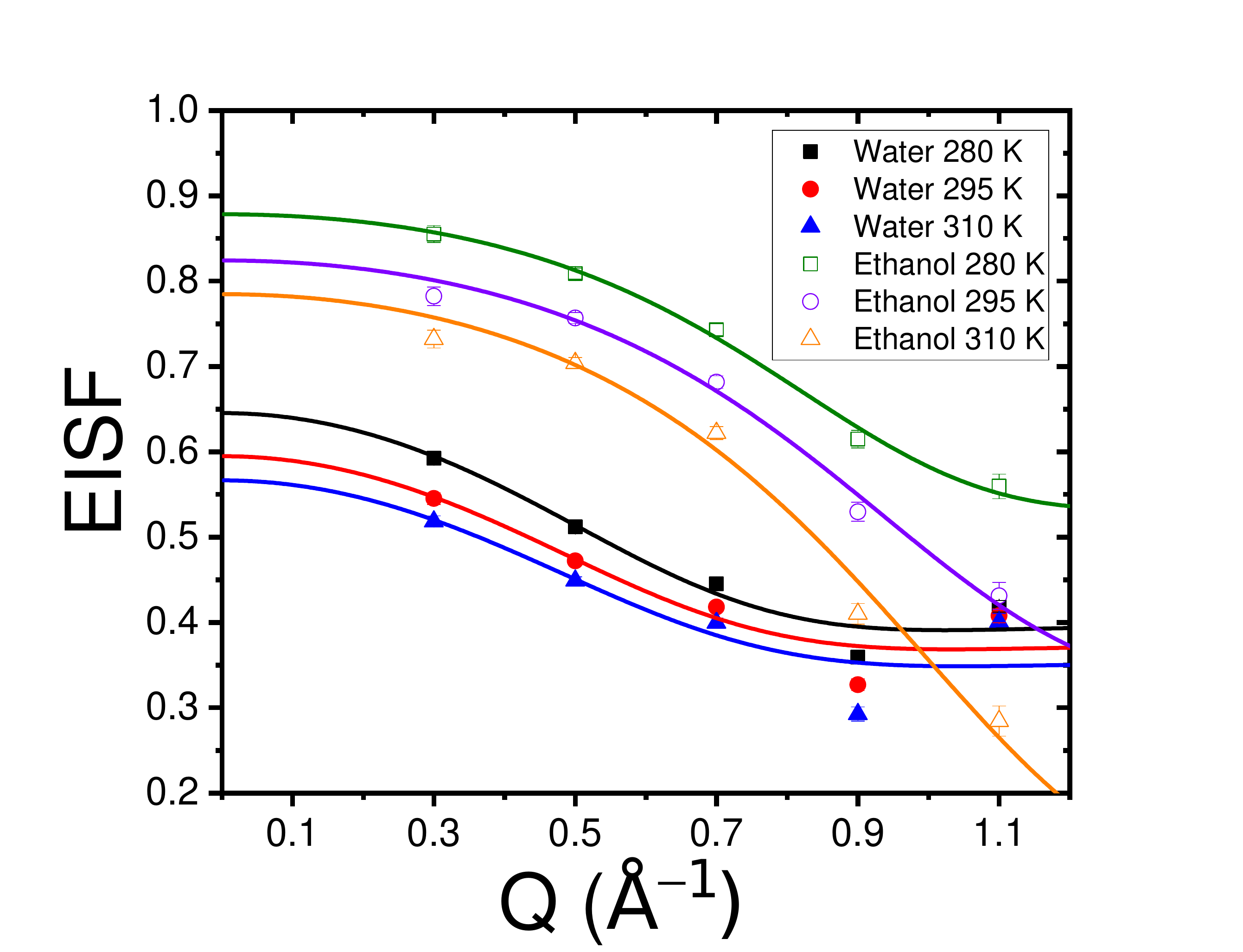}
   
    \caption{Elastic Incoherent Structure Factor (EISF) for water and ethanol at various temperatures measured in GO powder.}
    \label{EISS1}
    \end{center}
\end{figure}

Using this together with equation (\ref{EISF_eqn}), we obtained the fits shown in Figure \ref{EISS1} with fitting parameters shown in Table \ref{table1}. These fits are much improved compared to using equation (\ref{EISF_eqn}) by itself without taking the mobile species into account. The fact that the fits do not go $1$ at low $Q$ is because the maximum EISS will opnly reach the fraction $f$ of immobile/confined species, which is the species giving rise to the elastic peak.

\begin{figure}
    \centering
    \includegraphics[width=8.5cm]{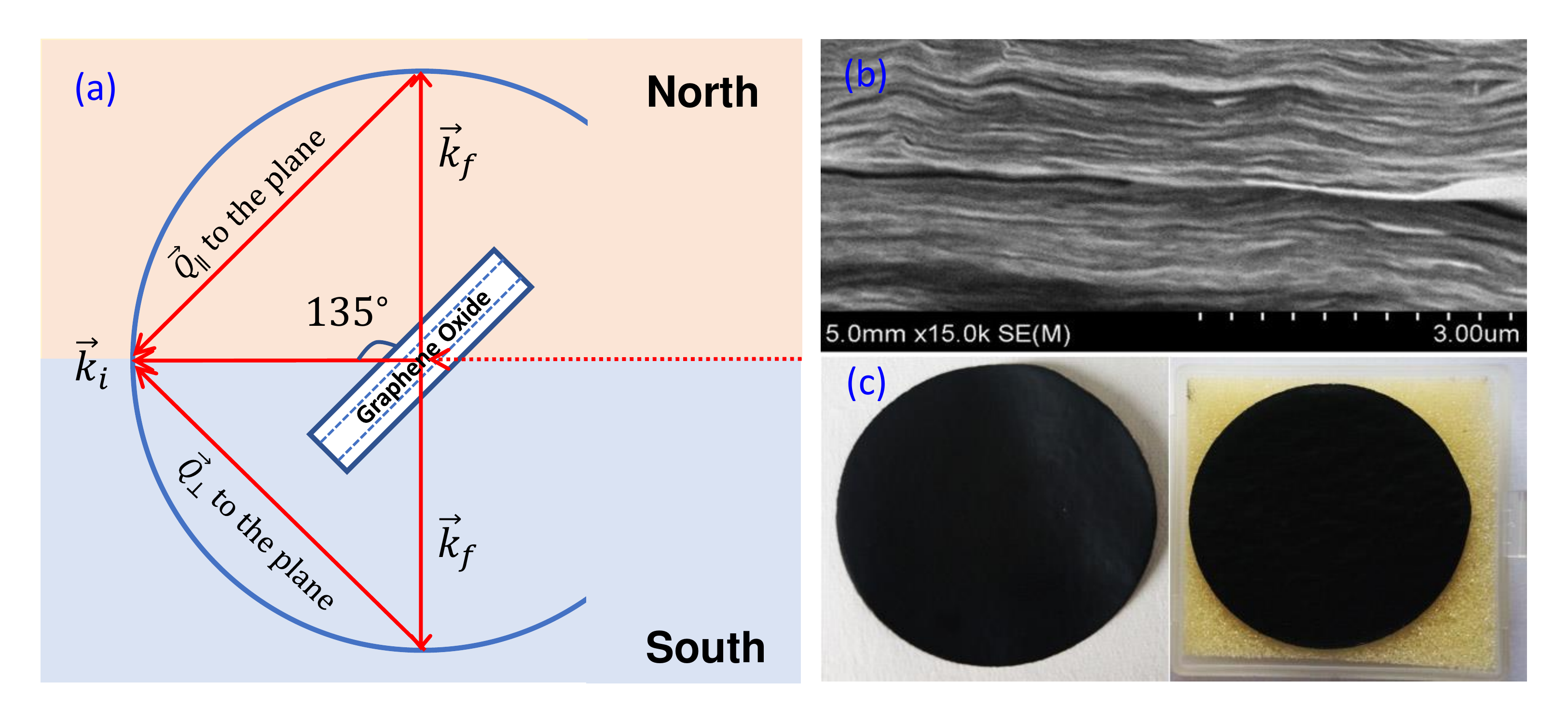}
    \caption{(a) BASIS geometry that allows probing anisotropic diffusion in oriented samples. (b) SEM image of (c) commercially available GO membrane. Images (b) and (c) were provided by kind permission of ACS Material LLC.}
    \label{sample_fig}
\end{figure}


Graphene oxide is believed to consist of a graphene-like carbon network, with epoxide and hydroxide groups in a random or semi-ordered pattern distributed on the carbon surface \cite{Mouhat2020}. The carbon-oxygen ratio in these system depends on the preparation method and values ranging from 2:1 to 20:1 have been reported, with most suggesting a typical range of 2-5:1 \cite{Mouhat2020}. The lattice constant of graphene is $2.46 \textrm{\AA}$, suggesting that epoxide or hydroxyl species are on average spaced $5-10 \textrm{\AA}$. This is consistent with the jump distances $l$ we found in our analysis. The values for $a$ are slightly smaller than $l$, which is not surprising. 

We therefore arrive at a picture where some molecules are immobilized to various degrees around the surface oxide species, while the remaining molecules within the channel diffuse by jump diffusion. This mobile species is retarded in their motion by the presence of the oxygen groups on the surface (but not immobilized on the time scale of the experiment), leading to longer relaxation times $\tau_0$ and jump distances that are consistent with the spacing of the epoxide/hydroxide groups.

Based on these results, what could be the reason for the different temperature dependence of the dynamics of water and ethanol? In water, we see an Arrhenius-like increase in diffusivity, while in ethanol we did not see an increase. This may be related to the an increase in the jump distance $l$ with temperature seen for water (but not for ethanol). 

In water, the fraction of completely immobilized $p_0$ molecules, as well as the effective confinement radius, remain constant with temperature. This suggests that these are bonded too tightly to be affected by thermal motion in the narrow temperature range we probed. There is also a hint of a slight decrease of confined/immobilized species($f$) with temperature. 

In ethanol, we see no change in $l$ with temperature, but a clear decrease in immobilized molecules ($p_0$). This clearly different behavior of water and ethanol could be because water hydrogen bonding to surface hydroxides is twice as strong as hydrogen bonding to ethanol \cite{Neklyudov2017}.

\begin{figure}[h]
    \begin{center}
    \includegraphics[width=9.0cm]{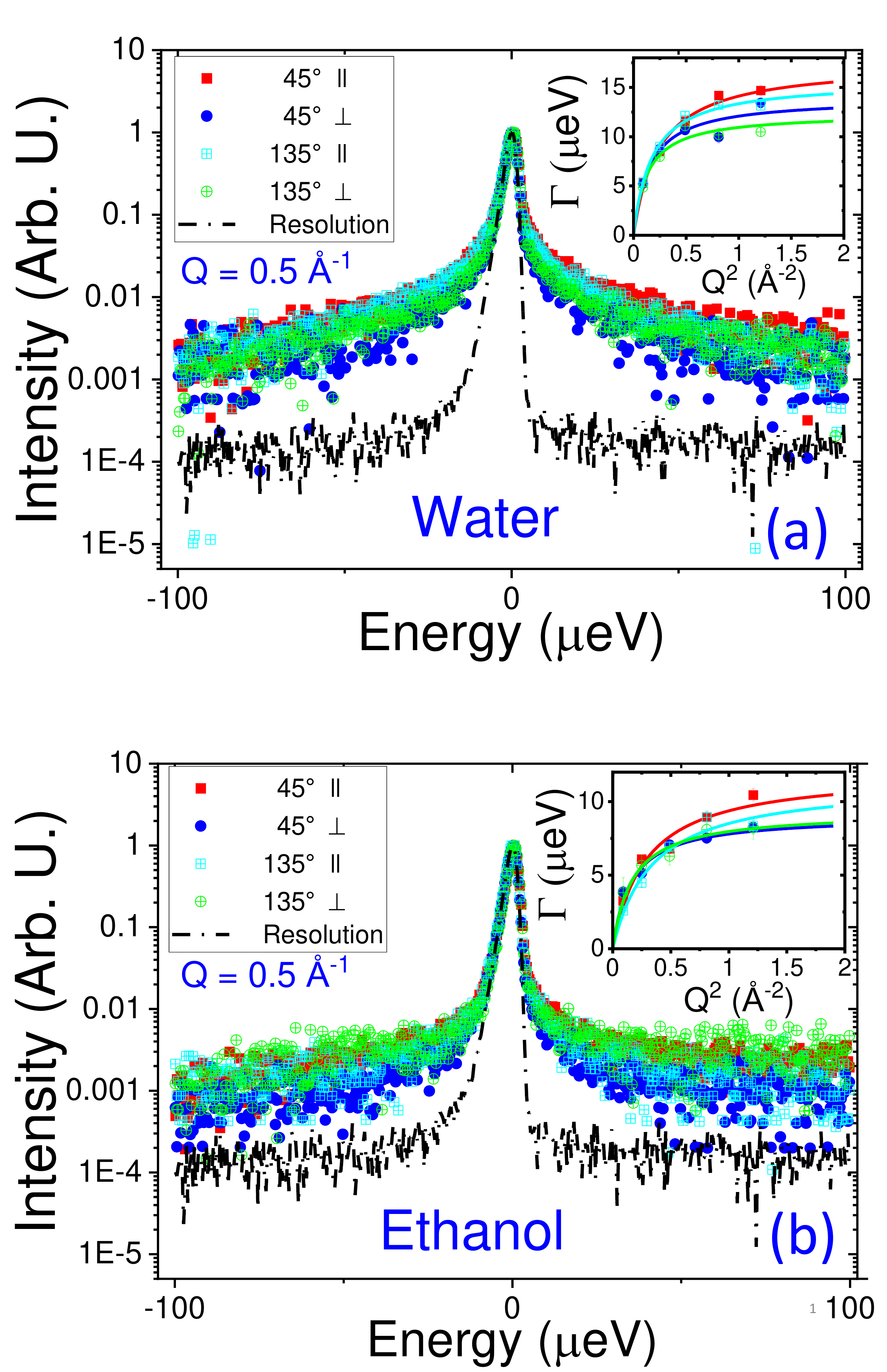}
    \caption{QENS Spectra from  water intercalated GO Membrane samples  at $Q = 0.5$ \AA $^{-1}$ and 310 K. The spectra were collected with $Q's$ in parallel ($\parallel$) and perpendicular ($\perp$) orientations as shown in Figure \ref{sample_fig}(a).}
    \label{QENS_Aniso}
    \end{center}
\end{figure}

\begin{figure}[h]
    \begin{center}
     \includegraphics[width=10.0cm]{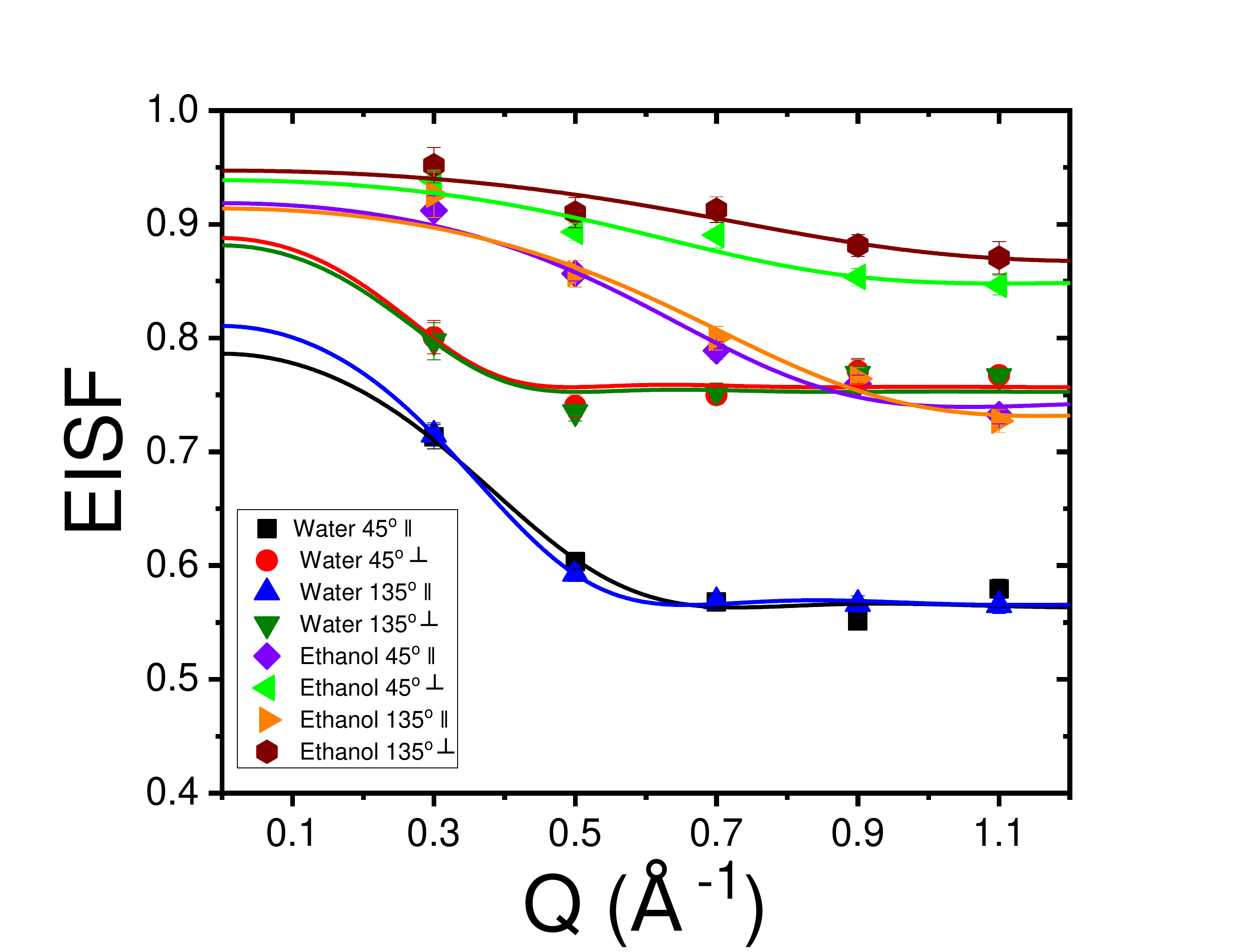}
    \caption{Elastic Incoherent Structure Factor (EISF) for water and ethanol in various orientations. The angle ($45^\circ$ or $135^\circ$) indicates the incident angle of the beam, while $\parallel$ and $\perp$ indicate diffusion parallel or perpendicular to the planes of the membrane. It can be seen that the results are not dependent on the incidence angle of the beam. Rather, the results show anisotropy with respect to direction of diffusion relative to the membrane.}
    \label{EISS2}
    \end{center}
\end{figure}

While the data we have discussed so far represent isotropically averaged diffusivities measured from GO powders, it is important to understand the anisotropy of solvents' diffusivity between the direction perpendicular to the membrane plane and the in-plane (parallel) direction. This may help to further understand the dynamics of the two molecular species in the system. We therefore utilized the Q-dependence of QENS signal to probe anisotropy of water and ethanol diffusivity in a graphene oxide membrane (Figure \ref{sample_fig}c), as shown in Figure \ref{QENS_Aniso}. 

Highly oriented graphene oxide membranes were used to measure QENS spectra at different orientations. QENS spectra for water and ethanol intercalated in GO membranes were measured in two different configurations: i) sample plane at $45^\textrm{o}$ and ii) sample plane at $135^\textrm{o}$ to the beam direction, as shown in Figure \ref{sample_fig}(a).  In this type of configuration, if the scattering vector on the north-side is perpendicular to the sample plane then it would be parallel to the south-side and vice versa. The obtained diffusivities are presented in table \ref{table2}.
\begin{table*}
\caption{\label{table2}%
Parameters obtained for water and ethanol intercalated in Graphene Oxide (membrane) at $310$ K at orientation shown in figure \ref{sample_fig}(a)}
\begin{ruledtabular}
\begin{tabular}{c c c c c c c c}
 
\textrm{Liquid }&
\textrm{Orientation}&
\textrm{D $\times$ 10$^{-10}$ m$^{-2}$s$^{-1}$}&
\textrm{ $\tau_0$ (ps)}&
\textrm{$l$ (\AA)}&
\textrm{f}&
\textrm{$p_0$}&
\textrm{$a$ (\AA)}
\\
\colrule
              &$45^o_{\parallel}$ & $11.08\pm1.34$ &$37.7\pm2.8 $ &$5.0\pm0.4$&$0.79\pm0.03$ & $0.35\pm 0.07$ & $6.2\pm0.6$\\
Water         &$45^o_{\perp}$ & $13.85\pm3.23$ &$47.3\pm4.2 $&$6.2\pm0.8$ &$0.89\pm0.57$ & $0.39\pm 2.29$ & $9.1\pm21.9$ \\
              &$135^o_{\parallel}$ & $13.66\pm1.19$ &$42.1\pm1.8 $&$5.9\pm0.3$ &$0.81\pm0.01$ & $0.30\pm 0.01$ & $6.8\pm0.1$ \\
              &$135^o_{\perp}$ & $14.52\pm3.70$ &$53.4\pm4.8 $&$6.8\pm0.9$  &$0.88\pm1.31$ & $0.40\pm 5.15$ & $8.9\pm51.0$  \\
    
          \hline
          
              &$45^o_{\parallel}$ & $6.41\pm1.56$ &$54.7\pm7.0 $&$4.6\pm0.6$  &$0.91\pm0.01$ & $0.25\pm 0.04$ & $4.3\pm0.4$  \\
Ethanol       &$45^o_{\perp}$ & $8.67\pm1.58$ &$73.0\pm4.9 $&$6.2\pm0.6$ &$0.93\pm0.01$ & $0.36\pm 0.09$ & $4.1\pm0.9$\\
              &$135^o_{\parallel}$ & $5.12\pm1.12$ &$57.9\pm6.9 $&$4.2\pm0.5$&$0.91\pm0.02$ & $0.25\pm 0.06$ & $3.9\pm0.5$ \\
           &$^{*}135^o_{\perp}$&  $8.42\pm2.29$ & $70.9\pm5.9$ & $6.0\pm0.8$&$0.95\pm0.02$ &$0.36\pm 0.10$ & $3.5\pm1.0$ \\     
\end{tabular}
\end{ruledtabular}
* The offsets of linear background for third and fifth Q's are constrained to obtain reasonable parameters while fitting QENS spectra.
\end{table*}
The data suggests slight differences between in--plane (parallel) and out-of-plane (perpendicular) diffusion. However, within the error bars, these differences were not significant (the most significant differences were found for the relaxation time and jump distances in ethanol with a $p_0$ value of about 0.1). Therefore, either diffusion is isotropic, or diffusion is slightly enhanced (faster) \emph{perpendicular} to the GO layers. Naively, we would expect that diffusion \emph{in-plane} is slightly faster in-plane than out-of-plane, although the difference is expected to be small \cite{aggarwal2007anisotropic}.

Diffusion \emph{through} the GO layers is expected to be about one order-of-magnitude slower than bulk diffusion \cite{devanathan2016molecular}. The most likely explanation for the measured data is therefore that diffusion is that we measure diffusion within layers of GO, rather than diffusion through the GO layers. The slight differences in jump distances $l$ between the parallel and the perpendicular direction are most likely geometry dependent. If molecules primarily jump between epoxide/hydroxide groups, these jump distances may reflect the average spacing between them either parallel or perpendicular to the membrane. This higher jump distance (at least for ethanol) may then explain the apparent higher diffusivity. If this picture is correct, then the confining walls seem to have little influence in the diffusion beyond the interactions with the hydrophilic epoxide/hydroxide groups. This would make sense, as graphene by itself is highly hydrophobic and should allow for fast diffusion \cite{Park_water_graphene2010}.

We also determined the EISF as a function of orientation (Figure \ref{EISS2} and Table \ref{table2}). This data also shows little anisotropy. The largest difference is in the $p_0$ values for ethanol, but even this difference is not statistically significant within the error bars.

\section{Conclusions}
Performing quasi-elastic neutron scattering measurements at BASIS (ORNL) and HFBS (NIST), we measured the diffusional dynamics of water and ethanol nano-confined in graphene oxide powders and membranes. We found that in both cases, the transition temperature from a molecular "frozen" phase to a mobile phase was depressed compared to bulk, although the effect was more dramatic for water ($\Delta T\approx 64 \textrm{K}$), than for ethanol ($\Delta T\approx 15 \textrm{K}$).

For BASIS, we determined the dynamic susceptibility and found that within the energy resolution of the instrument, we are able to capture a single relaxation process. Based on this, we fit the QENS data to a single Lorentzian, and found a Q dependence of the structure factor that suggests jump diffusion.  Performing measurements at varying temperatures, we found broadening of the QENS spectra and near-Arrhenius behavior for the diffusivity of water. However, in ethanol, we found no significant broadening and a near constant diffusivity over the temperature range we studied (280-310 K).

For both liquids, the measured diffusivities were less than their corresponding bulk values and, in the case of water, showed a reduction of about 30\% compared to bulk. We found that the activation barrier for diffusion is either similar or slightly reduced compared to bulk. The relaxation times we measured were long compared to single-molecular relaxation times. It should be kept in mind that the energy resolution of the instrument limited us to measuring relaxation times that are longer than about 7 ps. 

The different behavior of water versus ethanol as a function of temperature was further investigated by determining the elastic intensity structure factor (EISF). The presence of a significant elastic portion hints at the presence of an immobile (on the time scale of the experiment) or locally confined species in the system. We therefore postulate that there are two species of liquid molecules in the system: Immobile/locally confined species, likely immobilized around oxide/hydroxide groups, and a mobile species that diffuses by jump diffusion. The diffusion of the mobile species may be retarded by weak interactions with the epoxide/hydroxide groups, leading to weak confinement and jump-diffusional behavior. The measured jump distances are consistent with the expected average spacing of such surface groups.

The differences in temperature behavior of water versus ethanol may be related to the fact the hydrogen bonding between the solvent molecules and surface hydroxide groups is twice as strong for water than for ethanol. Therefore, the amount of immobilized ethanol may be more variable with temperature, while for water the immobilized water "cluster" may increase in effective size due to vibrations without a change in the fraction of immobilized water molecules.

By confining water or ethanol in oriented membranes, we probed the anisotropy of the dynamics of the confined liquids. QENS fits suggested a weak anisotropy in the dynamics of the diffusing species, which was however not statistically significant within the errors of the measurement. EISF data suggested little anisotropy of the dynamics of the immobile/locally confined species. Therefore, we conclude that either diffusion is isotropic or there is a weak anisotropy where diffusion is slightly faster in the perpendicular direction to the membrane. This may be related to  differences in parallel versus perpendicular spacing of the epoxide/hydroxide-groups. This weak anisotropy suggests that - except for the presence of epoxide/hydroxide groups, which lead to local confinement - the presence of the graphene walls has little effect on the dynamics of the confined liquids.

\section{Acknowledgements}
We acknowledge funding through Wayne State University, as well as beam time access through Oak Ridge National Laboratory (Spallation Neutron Source, SNS) and the National Institute of Standards and Technology, NIST Center for Neutron Research (NCNR). Work at ORNL’s Spallation Neutron Source was sponsored by the Scientific User Facilities Division, Office of Basic Energy Sciences, U.S. Department of Energy. Access to the HFBS was provided by the Center for High Resolution Neutron Scattering, a partnership between the National Institute of Standards and Technology and the National Science Foundation under Agreement No. DMR-2010792. The identification of any commercial product or trade name does not imply endorsement or
recommendation by NIST. Preliminary work on confined water was funded by NSF DMR-0804283.

\bibliography{my_bib_paper.bib}
\end{document}